\documentclass{IEEEtran4PSCC}
\ifCLASSINFOpdf
   \usepackage[pdftex]{graphicx}
   \DeclareGraphicsExtensions{.pdf,.jpeg,.png}
\else
\fi
\usepackage[cmex10]{amsmath}
\usepackage{bm}

\usepackage{caption} 
\usepackage{subcaption} 

\usepackage[draft]{minted}
\usepackage{float}
\floatstyle{ruled}
\newfloat{code}{thp}{lop}
\floatname{code}{Code Block}
\usepackage{color}

\begin{document}

\title{PowerModelsRestoration.jl: An Open-Source Framework for Exploring Power Network Restoration Algorithms}

\author{
\IEEEauthorblockN{Noah Rhodes$^1$, David M Fobes$^2$, Carleton Coffrin$^2$ and Line Roald$^1$}
\IEEEauthorblockA{$^1$University of Wisconsin-Madison,
Madison, Wisconsin, USA. Email:
\{nrhodes,roald\}@wisc.edu\\
$^2$Los Alamos National Laboratory, Los Alamos, New Mexico, USA. Email: \{dfobes,cjc\}@lanl.gov}
}

\maketitle

\begin{abstract}
With the escalating frequency of extreme grid disturbances, such as natural disasters, comes an increasing need for efficient recovery plans. Algorithms for optimal power restoration play an important role in developing such plans, but also give rise to challenging mixed-integer nonlinear optimization problems, where tractable solution methods are not yet available.  
To assist in research on such solution methods, this work proposes PowerModelsRestoration, a flexible, open-source software framework for rapidly designing and testing power restoration algorithms.  PowerModelsRestoration constructs a mathematical modeling layer for formalizing core restoration tasks that can be combined to develop complex workflows and high performance heuristics.  The efficacy of the proposed framework is demonstrated by proof-of-concept studies on three established cases from the literature, focusing on single-phase positive sequence network models. The results demonstrate that PowerModelsRestoration reproduces the established literature, and for the first time provide an analysis of restoration with nonlinear power flow models, which have not been previously considered.
\end{abstract}


\section{Introduction}
As the threat of exogenous grid disturbances, \textit{e.g.}, natural disasters or sophisticated targeted attacks, continues to intensify, so does the importance of expedient power network restoration.  There is thus an increasing need for decision support tools that can assist network operators in identifying optimal restoration plans, and ultimately provide autonomous self-healing capabilities to the grid.  Unfortunately, high-fidelity modeling of power system restoration has proven notoriously difficult due to the challenge of finding an AC-feasible operating point in networks where hundreds to thousands of components have been damaged \cite{Overbye:2004vb,6345338}.
However, recent advances in convex relaxations of the AC power flow equations have shown promise for  single-time-point N-k analysis \cite{coffrin2019}.

While traditional work on power system restoration has focused on system reenergization following a blackout where the vast majority of components are undamaged \cite{317561,adibi1987taskforce,kirschen1991guiding}, the work in this paper considers the longer-term problem of restoring power supply following extreme physical impacts, such as hurricanes or earth quakes, where a large number of components must be repaired before they can be re-energized.
Specifically, it focuses on component restoration ordering, \emph{i.e.}, selecting the sequence in which components should be prioritized for restoration, in order to minimize energy not served over time.  
This prioritization task results in a sequential network design problem, which, in practice, can be remarkably challenging to solve \cite{coffrin2012last}, due in part to the combination of discrete and continuous optimization variables as well as nonlinear, non-convex constraints.  Consequently, significant network approximations and/or heuristic methods are often required to make the problem tractable.
Optimizing the restoration order can provide a significant reduction in energy not served relative to what is achieved through applying simple rules or heuristics \cite{van2011vehicle}. 
However, previous work \cite{6345338,VanHentenryck2015} have demonstrated that popular approximations, \textit{e.g.}, DC power flow, fail to capture important aspects of the problem, leading to suboptimal restoration ordering, infeasible intermediate solutions, and higher-than-necessary energy not served.

Designing and validating the effectiveness of different approaches to this problem requires significant research, and is an essential step towards the aspiration of a resilient, self-healing grid.
To support this design challenge, the core contribution of this work is a novel software framework, PowerModelsRestoration, which enables rapid exploration of 
power network restoration algorithms.  By building on the PowerModels framework \cite{8442948}, PowerModelsRestoration is able to consider a broad range of power flow formulations, spanning the full AC equations \cite{Cain2012}, convex relaxations \cite{1664986,7271127}, and active-power-only approximations \cite{Stott_2009bb}.  This restoration framework includes exact restoration algorithms, modeled as mixed-integer nonlinear programs (MINLPs), and in the future will contain heuristic restoration algorithms, such as largest capability first.  Some of the notable features of PowerModelsRestoration include: (1) support for AC-based restoration, (2) restoration plan quality guarantees provided by convex relaxations, (3) incorporation of storage devices in power system restoration, and (4) tools for simulating restoration plans with the AC power flow equations.
With these features, PowerModelsRestoration aspires to be a valuable tool for rapidly exploring the wide variety of possible restoration algorithms, and providing a baseline implementation of established restoration algorithms for resilience analysis.

Utilizing the proposed software framework, this work develops case studies that highlight the benefits and drawbacks of established power flow formulations. On the one hand, the results demonstrate the flexibility and value of the software framework, while on the other hand provide new insights into restoration nonlinear formulations, not previously considered to our knowledge.  An unexpected contribution of the paper is the insight that convex relaxations of the power flow equations may provide a valuable tool for balancing accuracy and performance when developing power restoration plans.

This work begins with a brief overview of the power system restoration context that motivates this work in Section \ref{sec:background}, followed by a review of mathematical programming and the PowerModels framework in Section \ref{sec:powermodels}.  The mathematical modeling layer of PowerModelsRestoration is presented in Section \ref{sec:mrsp-rop}. Section \ref{sec:using_pmr} illustrates how the modeling layer can be combined into more complex analysis workflows. Section \ref{sec:study} develops restoration studies to validate the framework and Section \ref{sec:conclusion} concludes the paper.

\section{Power Restoration Background and Modelling}
\label{sec:background}

This section provides an overview of the main modeling concepts that currently form the basis of PowerModelsRestroration.

\subsection{Restoration of a physically damaged grid}
As noted above, traditionally in power system restoration work, it is assumed that the majority of components are still physically intact. 
For transmission grid restoration, the main objective of traditional methods is therefore to create a sequence in which components can be re-energized, considering, \textit{e.g.}, the black-start capabilities of power plants and the dynamics of cold-load pick-up and power plant re-synchronization \cite{8326544}. Distribution system restoration has focused on a problem called Power Supply Restoration (PSR) \cite{thiebaux2013planning,hijazi2014optimalreconfig,li2014spanningtree}, which is an operational task that 
considers how to reconfigure a distribution system's topology for the purposes of fault isolation and power resupply to a maximum possible amount of loads.

In contrast, we consider a different time scale, where the initial restoration and re-synchronization of available generators has already taken place, but a large number of system components are still damaged, and require physical repairs. 
This problem has been the topic of transmission system restoration research for many years \cite{317561} and forms the basis for the software implementation discussed in this paper.
Related work for distribution grids includes \cite{tan2019postdisaster}, which utilizes a connectivity-based model of the power network.

\subsection{Modelling assumptions}
In our current formulation, we assume: (1) a single-phase equivalent network representation; (2) full awareness of the system's state, including energized and damaged components; (3) remote controllability of all components, including, \emph{e.g.}, generator outputs and breaker setpoints; and (4) the ability to partially shed loads at buses.
In the context of a transmission system, where instruments such as measurement units, remote control devices, and circuit breakers are typically available throughout the system, these assumptions are reasonable. Additionally, transmission loads usually represent aggregations of many smaller loads, which can be switched via substation reconfiguration procedures, enabling gradual load shedding.

For distribution grids, these assumptions represent a simplification, which are often considered non-realistic.
In distribution grids the number of measurement devices is typically limited and information about damaged items requires manual surveying. There are typically fewer switches and breakers, so branches cannot be re-energized before several of them have been repaired, and loads are often discrete. 
Finally, distribution systems are usually unbalanced (especially so during restoration), so a three-phase representation would be more appropriate. However, given the limited amount of previous work on distribution grids, which did not directly account for power flow or voltage magnitude constraints (but has considered limited number of branches \cite{li2014spanningtree}), we believe that modeling framework demonstrated here can also provide interesting insights into distribution restoration. Further details on how we plan to extend this framework to more realistic distribution grid models is provided in the future work section.

\subsection{Problem formulations}
Based on the above assumptions, we present two problems essential to both transmission and distribution grid restoration. 

\paragraph*{Minimum Restoration Set Problem (MRSP)}
In the interest of fault resiliency and routine maintenance, power systems are engineered with many redundant components.  Consequently, only a subset of the damaged components may need to be restored to meet the needs of all of the loads in the system.  The goal of the Minimum Restoration Set Problem (MRSP) is to identify the smallest set of components that must be repaired to support the entire system load.  
From an optimization perspective, the MRSP is a network design task, similar to Transmission Network Expansion Planning \cite{989203}, but tailored to the power restoration context. 

This problem includes discrete variables for each damaged component indicating whether the component will be repaired or remain out-of-service. To model whether a choice of repairs is feasible, the problem includes a set of equality constraints representing the power flow throughout the grid, as well as constraints on generation capacity, transmission line ratings, and voltage magnitudes for all in-service components. The problem focuses only on a single time point (\emph{e.g.}, near-term peak demand). A more detailed discussion on the MRSP problem formulation can be found in \cite{van2011vehicle}.

\paragraph*{Restoration Ordering Problem (ROP)}
The Restoration Ordering Problem (ROP) represents the most fundamental power system restoration optimization task.
This problem most closely reflects the goal of network operators during power restoration, \textit{i.e.}, to minimize the amount of load shed,
given a limited amount of resources. The objective of the ROP is to identify the sequence in which the components should be restored, while minimizing unserved load over time, typically referred to as \emph{energy not served (ENS)}. From an optimization perspective, the ROP is a generalization of the MRSP, a so-called sequential network design task \cite{van2011vehicle,KALINOWSKI201551}.

The ROP problem is divided into finite time steps, and in each time step a limited number of components can be repaired (representing resource constraints on, \textit{e.g.}, the number of repair teams). After a component has been repaired, it can either immediately be put into service, or wait for other components to be restored. In our model, we only include variables to indicate whether a component is out-of-service (damaged, or repaired but not yet energized) or in-service (not damaged, or repaired and energized).
After a component has been energized, it must remain in service for the remainder of the restoration. Similar to the MRSP, at each step the ROP includes a model of both the power flow and the generation and transmission constraints. Our current implementation also assumes a constant load profile, representing the daily peak load (assuming restoration takes days or weeks to be completed).

\subsection{Computational challenges}
The MRSP and ROP tasks include both continuous and discrete decision variables. Such problems present significant computational challenges, particularly when combined with detailed representations of the power flow, such as the nonlinear, non-convex AC power flow model.
Particularly the ROP problem, which includes temporal constraints linking the restoration sequence, is very challenging to solve in practice. 

Current optimization solvers are inadequate for reliably solving the MINLP presented by AC power system restoration on a large scale \cite{coffrin2019}. A common approach is to optimize the restoration problems with approximate linear power flow models (\emph{e.g.}, DC Power Flow), which can utilize outstanding commercial optimization tools, such as Gurobi \cite{gurobi} and CPLEX \cite{cplex}.  However, this approach requires the conversion of restoration plans into AC-feasible plans, proven to be significantly challenging \cite{6345338,COFFRIN2015144}.  Fortunately, recent works have proposed new promising approximations \cite{LPAC_ijoc} and relaxations \cite{coffrin2019} to address these shortcomings. PowerModelsRestoration brings these approximations and relaxations together within a common framework, convenient for exploring multiple formulations in terms of accuracy and computational efficiency.

\section{PowerModelsRestoration}
\label{sec:powermodels}
Although many approximations or relaxations of the power flow equations \cite{coffrin2019} have been proposed, it is unclear which are suitable for power system restoration \cite{6345338,VanHentenryck2015}.
To address this, it is necessary to (1) develop several implementations of power system restoration models with differing levels of modeling accuracy for performance analysis and model validation, and (2) devise workflows that decompose complex power restoration tasks into multiple subproblems that can be solved effectively at scale \cite{van2011vehicle,coffrin2012last}. A central goal of this work is to support researchers in addressing these challenges.

\subsection{PowerModels}
The PowerModels framework \cite{8442948}, which was designed to streamline model development and assist in exploring the accuracy and computational efficiency of different power flow formulations, is a core inspiration for this work. 
PowerModels is implemented in the mathematical programming framework JuMP \cite{DunningHuchetteLubin2017}, which has emerged as a commercial-grade tool for optimization modeling in Julia \cite{julia}, a high-performance programming language for numerical computing.  PowerModels \cite{8442948} was created as a modeling layer specializing in mathematical programs for power systems focusing on enabling easy switching between a variety of power flow formulations, which has been particularly useful for research on efficient formulations of AC Optimal Power Flow (OPF) \cite{7271127}.

PowerModels allows the development of formulation-agnostic power system optimization models using power-network-aware abstractions.  These abstractions are transformed into concrete mathematical models upon specification of a formulation. An extensive list of power flow formulations is provided with the PowerModels documentation, but below we illustrate a simple example of how these abstractions work.

\paragraph*{Example of PowerModels Abstractions} 
A power network optimization task, \textit{e.g.}, OPF, involves several expressions that include the complex voltage product, $V_i V^*_j$.  Depending on the formulation specified by the user, PowerModels replaces these generic expressions with a specific real number implementation.  For example, specifying the {\em ACPPowerModel} formulation (the AC power flow formulation in polar coordinates) results in the following mapping:
\begin{align}
V_i V^*_j &\Rightarrow |V_i||V_j|\cos(\theta_i - \theta_j) + \bm i |V_i||V_j|\sin(\theta_i - \theta_j)
\end{align}
If the {\em DCPPowerModel} formulation (the traditional DC power flow approximation) is specified, the mapping is,
\begin{align}
V_i V^*_j &\Rightarrow 1.0 + \bm i (\theta_i - \theta_j)
\end{align}
while the {\em SOCWRPowerModel} relaxation from \cite{1664986} results in,
\begin{subequations}
\begin{align}
& V_i V^*_j \Rightarrow W^R_{ij} + \bm i W^I_{ij} \\
& (W^R_{ij})^2 + (W^I_{ij})^2 \leq W^R_{ii}W^R_{jj}
\end{align}
\end{subequations}
For more general expressions, this mapping may be complex and can add many auxiliary variables and constraints.  

\subsection{Extension to PowerModelsRestoration}

This work introduces PowerModelsRestoration, a framework for studying power system restoration optimization. As an extension of PowerModels, 
PowerModelsRestoration can support a variety of power flow formulations, including AC Power Flow (ACPPowerModel), DC Power Flow (DCPPowerModel), and the SOC convex relaxation (SOCWRPowerModel) \cite{1664986}.  Furthermore, its flexible design similar to PowerModels enables and encourages user-driven extension. 
For example, while here we present a single-phase equivalent reformulation of the network, development to extend the framework to support multi-phase distribution network restoration via PowerModelsDistribution is underway.

\subsection{Restoration Optimization Specifications}
\label{sec:mrsp-rop}
In the following, we provide working examples of the two canonical restoration problems introduced above, the MRSP and ROP.
Detailed derivations of the mathematical models for each problem are available in \cite{van2011vehicle,VanHentenryck2015}, and
a more complete description of all constraints and variables can be found in software documentation for PowerModels\footnote{https://lanl-ansi.github.io/PowerModels.jl/stable} and PowerModelsRestoration\footnote{https://lanl-ansi.github.io/PowerModelsRestoration.jl/stable}. 
The goal of our presentation here is to demonstrate how the PowerModelsRestoration modeling layer is leveraged to implement these mathematical programs for power system restoration with a variety of power flow formulations. We also include functional code examples, which serve as concise illustrations of how the problems are constructed in PowerModelsRestoration v0.5.  Note that while PowerModelsRestoration includes variables and constraints for energy storage devices, we have omitted these components in the examples below for brevity and clarity.

\paragraph*{Minimum Restoration Set Problem (MRSP)}
We first discuss the implementation of the MRSP problem.
Code Block \ref{code:mrsp} shows the specification of the MRSP introduced informally in Section \ref{sec:background} as implemented in PowerModelsRestoration.  
First, optimization variables are added to the model, including all variables typically present in an OPF problem, which are continuous, as well as a discrete 0/1 variable representing the status of a component, indicated by the suffix \texttt{\_indicator}.
Next, constraints linking the optimization variables are added; cross referencing Code Block \ref{code:mrsp} with the OPF formulation in PowerModels \cite{8442948} highlights the model similarities, \emph{e.g.}, the re-use of the function \texttt{constraint\_power\_balance}, which represents the nodal power balance constraints. The key differences between the MRSP and standard OPF formulations are the \texttt{\_damage} constraints, which consider whether a component is active based on the status of its indicator variable.  Additionally, the constraints of the form \texttt{constraint\_{\em component}\_damage} implement the component dependency requirement that a bus must be active if anything connecting to it is active.
Finally, an objective function is added to minimize the number of restored components.

\begin{code}[t]
\caption{Abstract Minimum Restoration Set Model}
\label{code:mrsp}
\begin{minted}[numbersep=3pt,xleftmargin=8pt,linenos]{julia}
function build_mrsp(pm::AbstractPowerModel)
    # the mathematical program's optimization variables
    variable_bus_damage_indicator(pm)
    variable_bus_voltage_damage(pm)

    variable_branch_damage_indicator(pm)
    variable_branch_power(pm)

    variable_gen_damage_indicator(pm)
    variable_gen_power_damage(pm)

    # component by component power system constraints
    constraint_model_voltage_on_off(pm)

    for i in ids(pm, :ref_buses)
        constraint_theta_ref(pm, i)
    end

    for i in ids(pm, :bus)
        constraint_bus_damage_soft(pm, i)
        constraint_power_balance(pm, i)
    end

    for i in ids(pm, :gen)
        constraint_gen_damage(pm, i)
    end

    for i in ids(pm, :branch)
        constraint_branch_damage(pm, i)
        constraint_ohms_yt_from_damage(pm, i)
        constraint_ohms_yt_to_damage(pm, i)

        constraint_voltage_angle_difference_damage(pm, i)

        constraint_thermal_limit_from_damage(pm, i)
        constraint_thermal_limit_to_damage(pm, i)
    end

    # a minimum restoration count objective
    objective_min_restoration(pm)
end
\end{minted}
\end{code}

\paragraph*{Restoration Ordering Problem (ROP)}
We next discuss the implementation of the multi-period ROP.
Code Block \ref{code:rop} shows a specification of the ROP introduced in Section \ref{sec:background}.  As previously mentioned, the ROP is a multi-time point generalization of the MRSP. The sequential nature of the ROP leverages PowerModels' {\em multinetwork} feature, which enables multiple networks with linking constraints to be optimized simultaneously.  In this specification, $N$ copies of the network are specified in the data model, each representing a single time-point network restoration state, which are linked together with inter-temporal constraints.  It is assumed that that $n=1$ represents the initial state, and all damaged components can be restored by $n=N$. The number of restoration periods $N-1$ is specified by the user, and the per period limit on the number of restored items (constant for all periods) is automatically chosen such that all items can be restored by the final period.

The first for-loop of Code Block \ref{code:rop} constructs the independent restoration models for each time period, reusing much of the modeling seen in Code Block \ref{code:mrsp}.  The notable differences are (1) new variables for power demands (loads) and bus shunts, which are required for partial power restoration in each period, and (2) a cardinality constraint, which limits the number of restoration actions in each period. 

Perhaps the most crucial constraints in the ROP specification are the inter-temporal constraints appearing in the second outer for-loop of Code Block \ref{code:rop}.  The \texttt{energized} constraints ensure that once a component is activated it remains active throughout the remainder of the restoration, while the \texttt{increasing} constraints ensure that once a load is served, it remains so for the remainder of the restoration. The \texttt{restore\_all\_items} constraint ensures that all initially damaged components are repaired and energized by the restoration completion.  Finally, the objective function aims to maximize the load delivered throughout the restoration process.

\begin{code}[!th]
\caption{Abstract Restoration Ordering Model}
\label{code:rop}
\begin{minted}[numbersep=3pt,xleftmargin=8pt,linenos]{julia}
function build_rop(pm::AbstractPowerModel)
    # add multiple copies of the restoration network
    for (n, network) in nws(pm) 
        variable_bus_damage_indicator(pm, nw=n)
        variable_bus_voltage_damage(pm, nw=n)

        variable_branch_damage_indicator(pm, nw=n)
        variable_branch_power(pm, nw=n)

        variable_gen_damage_indicator(pm, nw=n)
        variable_gen_power_damage(pm, nw=n)

        variable_load_power_factor(pm, nw=n, relax=true)
        variable_shunt_admittance_factor(pm, nw=n,
            relax=true)

        constraint_restoration_cardinality_ub(pm, nw=n)

        constraint_model_voltage_damage(pm, nw=n)

        for i in ids(pm, :ref_buses, nw=n)
            constraint_theta_ref(pm, i, nw=n)
        end

        for i in ids(pm, :bus, nw=n)
            constraint_bus_damage_soft(pm, i, nw=n)
            constraint_power_balance_shed(pm, i, nw=n)
        end

        for i in ids(pm, :gen, nw=n)
            constraint_gen_damage(pm, i, nw=n)
        end

        for i in ids(pm, :load, nw=n)
            constraint_load_damage(pm, i, nw=n)
        end

        for i in ids(pm, :shunt, nw=n)
            constraint_shunt_damage(pm, i, nw=n)
        end

        for i in ids(pm, :branch, nw=n)
            constraint_branch_damage(pm, i, nw=n)
            constraint_ohms_yt_from_damage(pm, i, nw=n)
            constraint_ohms_yt_to_damage(pm, i, nw=n)

            constraint_voltage_angle_difference_damage(pm,
                i, nw=n)

            constraint_thermal_limit_from_damage(pm, i,
                nw=n)
            constraint_thermal_limit_to_damage(pm, i, nw=n)
        end
    end

    # add inter-temporal constraints across the networks
    network_ids = sort(collect(nw_ids(pm)))
    n_1 = network_ids[1]
    for n_2 in network_ids[2:end]
        for i in ids(pm, :gen, nw=n_2)
            constraint_gen_energized(pm, i, n_1, n_2)
        end
        for i in ids(pm, :bus, nw=n_2)
            constraint_bus_energized(pm, i, n_1, n_2)
        end
        for i in ids(pm, :branch, nw=n_2)
            constraint_branch_energized(pm, i, n_1, n_2)
        end
        for i in ids(pm, :load, nw=n_2)
            constraint_load_increasing(pm, i, n_1, n_2)
        end
        n_1 = n_2
    end

    n_final = last(network_ids)
    constraint_restore_all_items(pm, n_final)

    objective_max_load_delivered(pm)
end
\end{minted}
\end{code}

\section{Using PowerModelsRestoration}
\label{sec:using_pmr}
In this section we demonstrate how the mathematical programs for power system restoration highlighted in the previous section can be leveraged in simple Julia scripts to perform more complex power restoration analysis.  

\paragraph*{Choosing a solver} PowerModelsRestoration constructs a mathematical program (\emph{i.e.}, a JuMP model), which can be solved using a general purpose optimization solver. Many solvers are available in JuMP, including Ipopt for continuous nonlinear programs (NLP) \cite{Ipopt}, Cbc for mixed-integer linear programs (MIP) \cite{john_forrest_2018_1317566}, and Juniper for mixed-integer nonlinear programs (MINLP) \cite{juniper}.  Better performance may be gained by using commercial solvers, \textit{e.g.}, Gurobi or KNITRO.

\paragraph*{Restoration Optimization and Simulation}
A core workflow for power systems restoration is to compute a restoration plan using a relaxed or approximated power flow formulation and validate the resulting plan with a higher fidelity power flow representation \cite{6345338,VanHentenryck2015}. Different from the formulation in \cite{6345338,VanHentenryck2015}, which was based on running an AC power flow with the provided restoration sequence and generation setpoints, we implement a \emph{multi-period AC OPF} that only uses the ordering of the components to fix the binary variables values, and allow the redispatch to adjust the value of all other variables, such as choosing new generation setpoints.  
Code Block \ref{code:sim} provides an example of this workflow implemented in  PowerModelsRestoration. First, the DC power flow approximation is used to compute a restoration plan, and the resulting plan is then simulated via the inter-temporal AC maximal load delivery optimization (\emph{i.e.}, ACPPowerModel).

\begin{code}[t]
\caption{AC Redispatch of a DC Restoration Plan}
\label{code:sim}
\begin{minted}{julia}
using PowerModels, PowerModelsRestoration
using Ipopt, Cbc

# load the network data with component damage
case = PowerModels.parse_file("case5_strg_damaged.m")
restore_case = replicate_restoration_network(case, count=3)

# optimize a restoration plan using the DC model
restore_result = run_rop(restore_case, DCPPowerModel, 
    Cbc.Optimizer)

# update the data with the restoration plan
clean_status!(restore_result["solution"])
update_status!(restore_case, restore_result["solution"])

# simulate the restoration plan with the AC model
result = run_restoration_redispatch(restore_case,
    ACPPowerModel, Ipopt.Optimizer)
\end{minted}
\end{code}

\begin{code}[t]
\caption{Building a Three-Step AC Restoration Plan}
\label{code:mrsp_rop}
\begin{minted}{julia}
using PowerModels, PowerModelsRestoration, JuMP
using Ipopt, Juniper
ipopt = optimizer_with_attributes(Ipopt.Optimizer, 
    "print_level"=>0)
solver = optimizer_with_attributes(Juniper.Optimizer,
    "nl_solver"=>ipopt)

# load the network data with component damage
case = PowerModels.parse_file("case5_strg_damaged.m") 

# solve the minimum restoration set problem
result = run_mrsp(case, ACPPowerModel, solver) 

# update the case data with the minimum restoration set
clean_status!(result["solution"])
update_status!(case, result["solution"])

# compute a three-step restoration plan
restore_case = replicate_restoration_network(case, count=3)
restore_result = run_rop(restore_case, ACPPowerModel, solver)
\end{minted}
\end{code}

\paragraph*{Accelerating Restoration Optimization}
Another core workflow for improving the solve time of a large ROP instance is to first solve the MRSP and then solve the ROP including only the subset of damaged components identified by the MRSP solution \cite{van2011vehicle,VanHentenryck2015}. 
Code Block \ref{code:mrsp_rop} presents an example of this workflow. Communication between the MRSP and ROP models occurs in the network data specification, where the component {\em status} field is leveraged to indicate components that should be ignored in the ROP.  This example uses the AC power flow variant and specifies 3 time periods for the ROP.  Other power flow models can be used by changing the power flow specification to, \textit{e.g.}, SOCWRPowerModel.

\section{Case Study}
\label{sec:study}

The core features of PowerModelsRestoration are demonstrated by the following case study, using a small example for illustrative purposes, as well as two typical test cases representing both a transmission grid and a single-phase representation of a distribution grid. Beyond demonstrating the software, these simulations provide new insights into the impact of the different formulations and workflows. 
Throughout this section the following solvers were used: Ipopt v3.12 \cite{Wachter2006} for the NLP version of AC formulation, Gurobi v8.1 for the MIP and MISOCP models given by the DC and SOC formulations, and Juniper v0.6 \cite{Kroger} using Ipopt and Gurobi \cite{gurobi} as sub-solvers for the MINLP version of AC model.

\begin{figure}
    \begin{center}
    \includegraphics[width=0.65\columnwidth]{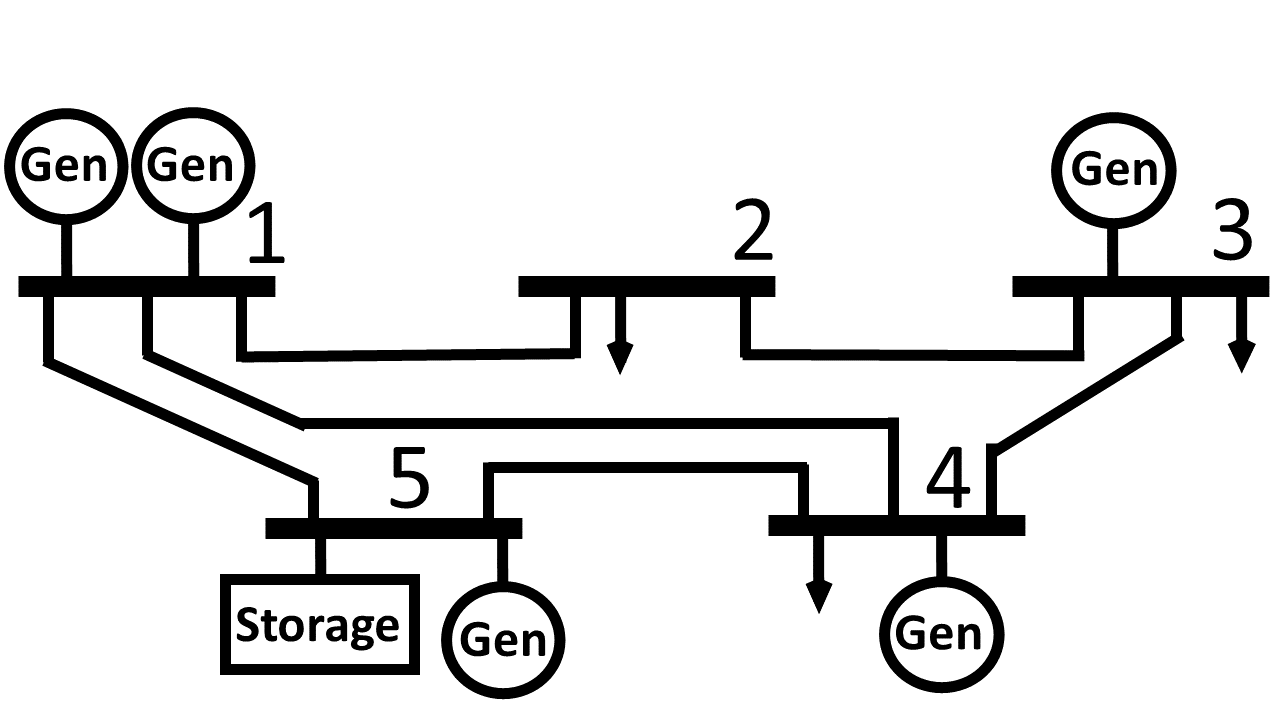}
    \end{center}
    \vspace{-0.3cm}
    \caption{\small A network diagram of a five-bus example illustrating core restoration decision problems. The generators on bus 1 have a small capacity, representing distributed energy resources. Note that bus 5 has energy storage.}
    \vspace{-0.2cm}
    \label{fig:5_bus}
\end{figure}

\subsection{Comparison of different power flow formulations}
The first case study investigates the the impact of power flow formulations on computational efficiency and quality of the solutions. For this investigation, we use a small 5-bus network \cite{li2010}, illustrated in Fig. \ref{fig:5_bus}. Using a small system allows for benchmarking simpler formulations against more detailed, but also more computationally intensive formulations. 
For the purposes of this study, all components in the network were damaged (except buses) and the workflow illustrated by code block \ref{code:sim} was utilized. 
We first solve the ROP problem using a DC, AC, and SOC power flow formulation, respectively. Each solution provides us with an optimal restoration sequence (from the optimal value of the decision variables), as well as an \emph{estimated ENS} (given by the objective function value). 
For each solution, we then solve a multi-step OPF using the full AC formulation to redispatch generators to obtain the \emph{true ENS}, except for the solution obtained with the AC ROP. For the AC ROP, the optimization already accounts for the full AC model and thus the estimated ENS would be the same as the true ENS.

We first compare the performance of different power flow formulations in terms of (i) the solve speed of the ROP problem, (ii) the estimated ENS, as obtained directly from the ROP, and (iii) the true ENS as computed by the AC redispatch (shown in Table \ref{table:model_comp}). We observe that the DC ROP is by far the fastest, requiring $\approx 0.1$s to reach an optimal solution, and the SOC ROP is also relatively quick, ($2.11$s). By comparison, the AC ROP requires more than $15$ minutes even for the 5-bus case, and can only to solve to local optimality which results in a solution with a comparatively higher ENS. 

\begin{figure*}[t]
    \centering
     \begin{subfigure}[b]{0.32\textwidth}
         \centering
         \includegraphics[width=0.97\textwidth]{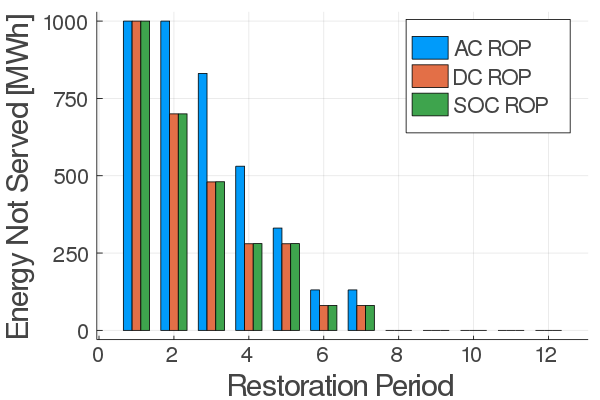}
         \caption{5-bus with different formulations}
         \label{fig:ENS_Formulations}
     \end{subfigure}
     \begin{subfigure}[b]{0.32\textwidth}
         \centering
        \includegraphics[width=0.98\textwidth]{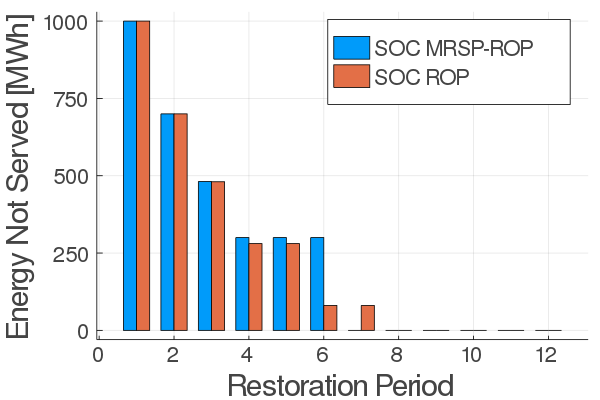}
         \caption{5-bus with MRSP preprocessing.}
         \label{fig:ENS_MRSP}
     \end{subfigure}    
     \begin{subfigure}[b]{0.32\textwidth}
         \centering
        \includegraphics[width=0.99\textwidth]{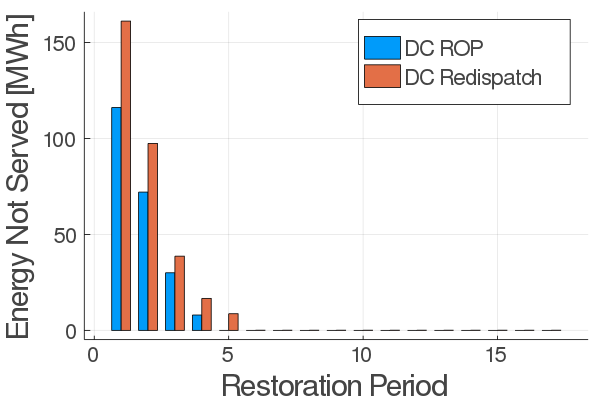}
         \caption{118-bus with DC formulation.}
         \label{fig:ENS_118}
     \end{subfigure}      
    \caption{\small (a) True Energy Not Served (ENS) for the 5-bus case with the AC, SOC and DC models. ENS is obtained from restoration redispatch using the full AC formulation. (b) True ENS for the ROP and MRSP+ROP problems using the SOC formulation in the fully damaged 5-bus case. (c) Estimated and true ENS per restoration period obtained from the DC ROP solution and corresponding AC resdispatch.}
\end{figure*}

In terms of solution quality, we first observe that using the AC power flow formulation provides inferior results, despite a higher fidelity model. This is due to higher problem complexity; the problem is non-convex even after the binary variables are fixed, and only a locally optimal solution can be found. 
Furthermore, we observe that the DC ROP returns the lowest estimated ENS. However, by comparing the estimated and true ENS, we note that the DC ROP problem is overly optimistic, and has a higher ENS (i.e. serves less load) than predicted. A similiar observation can be made for the SOC problem, although the gap between the estimated and true ENS is notably smaller. This indicates that the SOC ROP provides more accurate results than the DC ROP solution, with relatively small sacrifices in solve time. However, in this particular case, both the SOC and DC find a similar restoration sequence and are able to supply the same amount of load in the simulation.
To provide further details about the solutions, the ENS for each time step of the AC solutions is presented in Figure \ref{fig:ENS_Formulations}.  This plot highlights that the AC solution has a higher ENS in some of the time steps, and that the DC and SOC restoration sequences have very similar ENS in each period.

\begin{table}
    \centering 
    \caption{Comparison of power flow formulations (5-bus)}
    \label{table:model_comp}
    \begin{tabular}{r | c c c c}
        \hline
         Power Flow & Estimated & True & Solve   \\
         Formulation &  ENS [MWh]&  ENS [MWh]&  Time [s] \\
         \hline
         DC  & 2900 & 2902.5 & 0.12 \\
         SOC & 2902.4 & 2902.5 & 2.11 \\
         AC &  3953.9 & - & 914.72 \\
         \hline
    \end{tabular}
\end{table}

\subsection{MRSP as a preprocessing step}
The second case study considers the performance gains of applying MRSP as a preprocessing step, as well as the impact on solution quality.  
For this study, we use the same fully damaged 5-bus test case as above and the SOC power flow formulation. We compare two versions of the ROP problem, the original (as in the above section), and one where we apply MRSP as a preprocessing step to reduce the number of considered elements (\textit{i.e.}, the MRSP+ROP problem). This workflow is similar to the example shown in code block \ref{code:mrsp_rop}, and the results are shown in Table \ref{table:MRSP}.

The MRSP+ROP problem returns a solution in $ 0.103$ seconds (including the solve time for both the MRSP and the ROP problems), which is $\sim20\times$ faster than directly solving the ROP problem. This speed-up can be achieved because the MRSP problem reduces the number of components considered for repair from 11 to 6, meaning fewer restoration periods and fewer binary variables per time step in the ROP problem. We expect this speed enhancement to be more significant for larger systems containing more damaged components. 

In terms of solution quality, the MRSP+ROP produces a result with an ENS of 3081.4 MWh, notably higher than the restoration sequence from the full ROP (2902.4 MWh). To illustrate why this happens, we have plotted the ENS per time step for both the ROP and MRSP+ROP in Figure \ref{fig:ENS_MRSP}. It can be observed that by repairing only the minimum restoration set, the MRSP+ROP problem reaches full load delivery (\textit{i.e.}, ENS = 0) before the ROP solution. However, while the ROP solution repairs a larger number of components than is strictly necessary before reaching full load delivery, it serves more load {\em while the repairs are ongoing}.

\begin{table}
    \centering 
    \caption{\small Performance of SOC Power Flow Formulation with the MRSP pre-processor on the 5-bus case}
    \label{table:MRSP}
    \begin{tabular}{r | c c}
        \hline
         Workflow & ENS &  Time [s]\\
         \hline
        ROP      & 2902.4   & 2.11 \\
        MRSP+ROP &  3081.4  & 0.103 \\
         \hline
    \end{tabular}
\end{table}

\subsection{Larger restoration problems with DC power flow}

The third case study investigates the efficiency of the DC power flow formulation in larger networks. For these investigations, we use the IEEE 118-bus transmission case \cite{1908.02788} and a single-phase equivalent (SPE) representation of the IEEE 123-bus distribution feeder case\footnote{The single-phase representation contains only 56 buses, but we will still refer to the case as the 123-bus case.}, obtained from \cite{bolognani2016}. 
In the 118-bus case, we create a damage scenario by simulating significant localized damage to a third of the network, representing, \textit{e.g.}, an event such as a hurricane or an earthquake that causes severe damage to only one area of the transmission network.  The damage scenario is generated randomly with 35\% to branches and generators in Area 1 of the network, which includes buses 1--23, 25--32, 113--115, 
and 117.  In the 123-bus distribution feeder, we simulate the impact of a windstorm by randomly damaging 25\% of the branches in the network.
For these two cases, we apply the DC ROP with subsequent AC redispatch. We assume an ability to repair multiple elements per period, such that we have 10 available repair periods for the 118-bus case, and 15 periods for the 123-bus case. At the end of the simulation, we record the solution time, estimated ENS and true ENS (summarized in Table \ref{table:dc_large}).

\begin{table}
    \centering 
    \caption{\small Performance of DC Power Flow Formulation for the 118- and 123-bus cases}
    \label{table:dc_large}
    \begin{tabular}{r | c c c c}
        \hline
         Test& System &Estimated & True & Solve   \\
         Case & Type & ENS [MWh]&  ENS [MWh]&  Time [s] \\
         \hline
         118 bus & Transmission & 226.2 & 323.6 & 6.81 \\
         123 bus & Distribution SPE & 1395.5 & 1396.4 & 0.955 \\
         \hline
    \end{tabular}
\end{table}

Even for these larger test cases, the DC ROP problem solves within a few seconds, demonstrating the tractability of DC power flow based restoration algorithms. However, analysis of the solution quality reveals some important differences between the two test cases. For the IEEE 118-bus transmission case, there is a large discrepancy between the estimated and true ENS, indicating that the DC ROP drastically overestimates the amount of load that can be delivered. In contrast, for the IEEE 123-bus distribution feeder case, the estimated and true ENS are nearly equivalent. 

To understand these differences, we first plot the estimated and true ENS for the 118-bus case in Figure \ref{fig:ENS_118}. We observe that the true ENS is higher than the predicted ENS for all time periods. This can be explained by considering the structure of the network.  The 118-bus case represents a well connected network with 35 synchronous condensers. In this network, reactive power plays a significant role in ensuring feasibility, which is not recognized by the DC power flow formulation. Instead, the DC power flow formulation delays the repair of the synchronous condensers because they are unnecessary for a feasible DC solution.

We do not plot the results for the 123-bus case, because there is almost no difference between the estimated and true ENS in any of the time periods. Again, this can be explained by the structure of the case;
the distribution system in the IEEE 123-bus case is a radial network where the substation is the only source of both active and reactive power. Due to the radial network structure and other network parameters, the restoration problem focuses on connectivity over the details of power flow, which is captured equally well by the DC and AC models.
The estimated and true ENS therefore match, since the ROP prioritizes feeder branches to reconnect to the substation first. However, it remains to be demonstrated how close the obtained solution is to the true optimal solution in this case. While the single phase equivalent representation and the assumptions regarding the restoration process are less not entirely realistic for a distribution network, our results demonstrate how the network topology plays a role in whether or not a power flow formulation provides accurate results.

\section{Conclusion}
\label{sec:conclusion}

This work highlights the need for a new generation of power restoration algorithms to support grid resiliency in the context of large-scale disturbances,
and illustrates the formidable computational challenges faced by such algorithms.  To that end, PowerModelsRestoration is proposed as a flexible framework for the rapid exploration of power restoration algorithms.  A validation of the framework was conducted on seminal network cases, including a 5-bus test case, the IEEE 118-bus transmission network and the 123-bus distribution feeder network.  This validation replicated key results from related works \cite{van2011vehicle,VanHentenryck2015}, but show that the DC model is able to provide feasible (though potentially suboptimal) restoration sequences when generators are redispatched in a multi-period AC OPF framework. The paper provides novel insights into the challenges of non-convex restoration formulations and the successes of convex relaxations, which had not been considered previously.  Overall, PowerModelsRestoration represents a strong foundation for continued research on power restoration algorithms and resiliency analysis.

One important future direction of this work is to improve the support for restoration in distribution grids.  While we have notably included support for distributed energy resources in PowerModelsRestoration,  distribution restoration requires a number of additional considerations, including sectionalizing, limited availability of switching devices and loads, which can no longer be assumed to be continuous variables, as well as the consideration of three-phase, unbalanced grids, which could by achieved through integration with the PowerModelsDistribution package.

Another important avenue for future work is the integration of metaheuristics, such as hierarchical or adaptive decomposition schemes \cite{coffrin2012last}, for scaling the restoration ordering problem to much larger problems sizes, \textit{e.g.}, those with thousands of buses.
Furthermore, our current formulation only considers the ordering of power component repair, while ignoring other important aspects such as optimal repair crew dispatching \cite{van2011vehicle}, and stability margin verification during the restoration process. As such, the work presented here represents a sub-problem to the overall challenge of power grid recovery. However, addressing the challenges of this foundational sub-problem will enable us to efficiently solve the more comprehensive and complex problems in the future.

\section{Acknowledgements}
This work was supported by funding from the U.S. Department of Energy’s (DOE) Office of Electricity as part of the CleanStart-DERMS project of the Grid Modernization Laboratory Consortium and by the MACSER project, funded by U.S. Department of Energy’s (DOE) Office of Science.

\bibliographystyle{IEEEtran}
\bibliography{references}

\begin{thebibliography}{10}
\providecommand{\url}[1]{#1}
\csname url@samestyle\endcsname
\providecommand{\newblock}{\relax}
\providecommand{\bibinfo}[2]{#2}
\providecommand{\BIBentrySTDinterwordspacing}{\spaceskip=0pt\relax}
\providecommand{\BIBentryALTinterwordstretchfactor}{4}
\providecommand{\BIBentryALTinterwordspacing}{\spaceskip=\fontdimen2\font plus
\BIBentryALTinterwordstretchfactor\fontdimen3\font minus
  \fontdimen4\font\relax}
\providecommand{\BIBforeignlanguage}[2]{{%
\expandafter\ifx\csname l@#1\endcsname\relax
\typeout{** WARNING: IEEEtran.bst: No hyphenation pattern has been}%
\typeout{** loaded for the language `#1'. Using the pattern for}%
\typeout{** the default language instead.}%
\else
\language=\csname l@#1\endcsname
\fi
#2}}
\providecommand{\BIBdecl}{\relax}
\BIBdecl

\bibitem{Overbye:2004vb}
T.~Overbye, X.~Cheng, and Y.~Sun, ``A comparison of the {AC} and {DC} power
  flow models for {LMP} calculations,'' in \emph{Proceedings of the 37th Annual
  Hawaii International Conference on System Sciences}, 2004, p. 9 pp.

\bibitem{6345338}
C.~Coffrin, P.~V. Hentenryck, and R.~Bent, ``Accurate load and generation
  scheduling for linearized dc models with contingencies,'' in \emph{2012 IEEE
  Power and Energy Society General Meeting}, July 2012, pp. 1--8.

\bibitem{coffrin2019}
C.~{Coffrin}, R.~{Bent}, B.~{Tasseff}, K.~{Sundar}, and S.~{Backhaus},
  ``Relaxations of ac maximal load delivery for severe contingency analysis,''
  \emph{IEEE Transactions on Power Systems}, vol.~34, no.~2, pp. 1450--1458,
  March 2019.

\bibitem{317561}
M.~M. {Adibi} and L.~H. {Fink}, ``Power system restoration planning,''
  \emph{IEEE Transactions on Power Systems}, vol.~9, no.~1, pp. 22--28, Feb
  1994.

\bibitem{adibi1987taskforce}
M.~{Adibi}, P.~{Clelland}, L.~{Fink}, H.~{Happ}, R.~{Kafka}, J.~{Raine},
  D.~{Scheurer}, and F.~{Trefny}, ``Power system restoration - a task force
  report,'' \emph{IEEE Transactions on Power Systems}, vol.~2, no.~2, pp.
  271--277, May 1987.

\bibitem{kirschen1991guiding}
D.~S. {Kirschen} and T.~L. {Volkmann}, ``Guiding a power system restoration
  with an expert system,'' \emph{IEEE Transactions on Power Systems}, vol.~6,
  no.~2, pp. 558--566, May 1991.

\bibitem{coffrin2012last}
C.~Coffrin, P.~Van~Hentenryck, and R.~Bent, ``Last-mile restoration for
  multiple interdependent infrastructures.'' in \emph{AAAI}, vol.~12, 2012, pp.
  455--463.

\bibitem{van2011vehicle}
P.~Van~Hentenryck, C.~Coffrin, R.~Bent \emph{et~al.}, ``Vehicle routing for the
  last mile of power system restoration,'' in \emph{Proceedings of the 17th
  Power Systems Computation Conference (PSCC’11), Stockholm, Sweden}.\hskip
  1em plus 0.5em minus 0.4em\relax Citeseer, 2011.

\bibitem{VanHentenryck2015}
\BIBentryALTinterwordspacing
P.~Van~Hentenryck and C.~Coffrin, ``Transmission system repair and
  restoration,'' \emph{Mathematical Programming}, vol. 151, no.~1, pp.
  347--373, Jun 2015. [Online]. Available:
  \url{https://doi.org/10.1007/s10107-015-0887-0}
\BIBentrySTDinterwordspacing

\bibitem{8442948}
C.~Coffrin, R.~Bent, K.~Sundar, Y.~Ng, and M.~Lubin, ``Powermodels.jl: An
  open-source framework for exploring power flow formulations,'' in \emph{2018
  Power Systems Computation Conference (PSCC)}, June 2018, pp. 1--8.

\bibitem{Cain2012}
\BIBentryALTinterwordspacing
M.~B. Cain, R.~P. {O' Neill}, and A.~Castillo, ``{History of optimal power flow
  and formulations},'' Federal Energy Regulatory Commission, Tech. Rep., 2012.
  [Online]. Available:
  \url{https://www.ferc.gov/industries/electric/indus-act/market-planning/opf-papers/acopf-1-history-formulation-testing.pdf}
\BIBentrySTDinterwordspacing

\bibitem{1664986}
R.~A. Jabr, ``Radial distribution load flow using conic programming,''
  \emph{IEEE Transactions on Power Systems}, vol.~21, no.~3, pp. 1458--1459,
  Aug 2006.

\bibitem{7271127}
C.~{Coffrin}, H.~L. {Hijazi}, and P.~{Van Hentenryck}, ``The qc relaxation: A
  theoretical and computational study on optimal power flow,'' \emph{IEEE
  Transactions on Power Systems}, vol.~31, no.~4, pp. 3008--3018, July 2016.

\bibitem{Stott_2009bb}
B.~Stott, J.~Jardim, and O.~Alsac, ``Dc power flow revisited,'' \emph{IEEE
  Transactions on Power Systems}, vol.~24, no.~3, pp. 1290--1300, 2009.

\bibitem{8326544}
L.~{Sun}, Z.~{Lin}, Y.~{Xu}, F.~{Wen}, C.~{Zhang}, and Y.~{Xue}, ``Optimal
  skeleton-network restoration considering generator start-up sequence and load
  pickup,'' \emph{IEEE Transactions on Smart Grid}, vol.~10, no.~3, pp.
  3174--3185, May 2019.

\bibitem{thiebaux2013planning}
S.~Thi{\'e}baux, C.~Coffrin, H.~Hijazi, and J.~Slaney, ``Planning with mip for
  supply restoration in power distribution systems,'' in \emph{Twenty-Third
  International Joint Conference on Artificial Intelligence}, 2013.

\bibitem{hijazi2014optimalreconfig}
H.~L. {Hijazi} and S.~{Thiébaux}, ``Optimal ac distribution systems
  reconfiguration,'' in \emph{2014 Power Systems Computation Conference}, Aug
  2014, pp. 1--7.

\bibitem{li2014spanningtree}
J.~{Li}, X.~{Ma}, C.~{Liu}, and K.~P. {Schneider}, ``Distribution system
  restoration with microgrids using spanning tree search,'' \emph{IEEE
  Transactions on Power Systems}, vol.~29, no.~6, pp. 3021--3029, Nov 2014.

\bibitem{tan2019postdisaster}
Y.~{Tan}, F.~{Qiu}, A.~K. {Das}, D.~S. {Kirschen}, P.~{Arabshahi}, and
  J.~{Wang}, ``Scheduling post-disaster repairs in electricity distribution
  networks,'' \emph{IEEE Transactions on Power Systems}, vol.~34, no.~4, pp.
  2611--2621, July 2019.

\bibitem{989203}
R.~{Romero}, A.~{Monticelli}, A.~{Garcia}, and S.~{Haffner}, ``Test systems and
  mathematical models for transmission network expansion planning,'' \emph{IEE
  Proceedings - Generation, Transmission and Distribution}, vol. 149, no.~1,
  pp. 27--36, Jan 2002.

\bibitem{KALINOWSKI201551}
\BIBentryALTinterwordspacing
T.~Kalinowski, D.~Matsypura, and M.~W. Savelsbergh, ``Incremental network
  design with maximum flows,'' \emph{European Journal of Operational Research},
  vol. 242, no.~1, pp. 51 -- 62, 2015. [Online]. Available:
  \url{http://www.sciencedirect.com/science/article/pii/S0377221714008078}
\BIBentrySTDinterwordspacing

\bibitem{gurobi}
{Gurobi Optimization, Inc.}, ``Gurobi optimizer reference manual,'' Published
  online at \url{http://www.gurobi.com}, 2014.

\bibitem{cplex}
I.~IBM, ``{IBM ILOG CPLEX Optimization Studio},''
  \url{http://www-01.ibm.com/software/commerce/optimization/cplex-optimizer/},
  2014.

\bibitem{COFFRIN2015144}
\BIBentryALTinterwordspacing
C.~Coffrin and P.~V. Hentenryck, ``Transmission system restoration with
  co-optimization of repairs, load pickups, and generation dispatch,''
  \emph{International Journal of Electrical Power \& Energy Systems}, vol.~72,
  pp. 144 -- 154, 2015, the Special Issue for 18th Power Systems Computation
  Conference. [Online]. Available:
  \url{http://www.sciencedirect.com/science/article/pii/S0142061515001106}
\BIBentrySTDinterwordspacing

\bibitem{LPAC_ijoc}
\BIBentryALTinterwordspacing
C.~Coffrin and P.~Van~Hentenryck, ``A linear-programming approximation of ac
  power flows,'' \emph{INFORMS Journal on Computing}, vol.~26, no.~4, pp.
  718--734, 2014. [Online]. Available:
  \url{http://dx.doi.org/10.1287/ijoc.2014.0594}
\BIBentrySTDinterwordspacing

\bibitem{DunningHuchetteLubin2017}
I.~Dunning, J.~Huchette, and M.~Lubin, ``Jump: A modeling language for
  mathematical optimization,'' \emph{SIAM Review}, vol.~59, no.~2, pp.
  295--320, 2017.

\bibitem{julia}
\BIBentryALTinterwordspacing
J.~Bezanson, A.~Edelman, S.~Karpinski, and V.~Shah, ``Julia: A fresh approach
  to numerical computing,'' \emph{SIAM Review}, vol.~59, no.~1, pp. 65--98,
  2017. [Online]. Available: \url{https://doi.org/10.1137/141000671}
\BIBentrySTDinterwordspacing

\bibitem{Ipopt}
A.~W{\"a}chter and L.~T. Biegler, ``On the implementation of a primal-dual
  interior point filter line search algorithm for large-scale nonlinear
  programming,'' \emph{Mathematical Programming}, vol. 106, no.~1, pp. 25--57,
  2006.

\bibitem{john_forrest_2018_1317566}
\BIBentryALTinterwordspacing
J.~Forrest, T.~Ralphs, S.~Vigerske, LouHafer, B.~Kristjansson, jpfasano,
  EdwinStraver, M.~Lubin, H.~G. Santos, rlougee, and M.~Saltzman,
  ``coin-or/cbc: Version 2.9.9,'' Jul. 2018. [Online]. Available:
  \url{https://doi.org/10.5281/zenodo.1317566}
\BIBentrySTDinterwordspacing

\bibitem{juniper}
O.~Kr{\"o}ger, C.~Coffrin, H.~Hijazi, and H.~Nagarajan, ``Juniper: An
  open-source nonlinear branch-and-bound solver in julia,'' in
  \emph{Integration of Constraint Programming, Artificial Intelligence, and
  Operations Research}, W.-J. van Hoeve, Ed.\hskip 1em plus 0.5em minus
  0.4em\relax Cham: Springer International Publishing, 2018, pp. 377--386.

\bibitem{Wachter2006}
A.~W{\"{a}}chter and L.~T. Biegler, ``{On the implementation of primal-dual
  interior point filter line search algorithm for large-scale nonlinear
  programming},'' \emph{Math. Prog.}, vol. 106, no.~1, pp. 25--57, 2006.

\bibitem{Kroger}
O.~Kr{\"{o}}ger, C.~Coffrin, H.~Hijazi, and H.~Nagarajan, ``{Juniper: an
  open-source nonlinear branch-and-bound solver in Julia},'' in
  \emph{Integration of Constraint Programming, Artificial Intelligence, and
  Operations Research}.\hskip 1em plus 0.5em minus 0.4em\relax Springer
  International Publishing, 2018, pp. 377----386.

\bibitem{li2010}
F.~{Li} and R.~{Bo}, ``Small test systems for power system economic studies,''
  in \emph{IEEE PES General Meeting}, July 2010, pp. 1--4.

\bibitem{1908.02788}
S.~Babaeinejadsarookolaee, A.~Birchfield, R.~D. Christie, C.~Coffrin,
  C.~DeMarco, R.~Diao, M.~Ferris, S.~Fliscounakis, S.~Greene, R.~Huang,
  C.~Josz, R.~Korab, B.~Lesieutre, J.~Maeght, D.~K. Molzahn, T.~J. Overbye,
  P.~Panciatici, B.~Park, J.~Snodgrass, and R.~Zimmerman, ``The power grid
  library for benchmarking ac optimal power flow algorithms,'' 2019.

\bibitem{bolognani2016}
S.~Bolognani and S.~Zampieri, ``On the existence and linear approximation of
  the power flow solution in power distribution networks,'' \emph{IEEE Trans.
  Power Syst.}, vol.~31, no.~1, pp. 163--172, Jan. 2016.

\end{thebibliography}
\end{document}